\documentclass [12pt]{article}
\usepackage {graphicx}
\begin{document}

\title{An example of physical system with hyperbolic attractor of Smale -- Williams 
type}

\author{S.~P.~Kuznetsov}

\maketitle

\begin{center}
\textit{Saratov Division of Institute of Radio-Engineering and Electronics, 
Russian Academy of Sciences, Zelenaya 38, Saratov, 410019, Russia}
\vspace{2mm}
\end{center}

\begin{abstract}
A simple and transparent example of a non-autonomous flow system, with 
hyperbolic strange attractor is suggested. The system is constructed on a 
basis of two coupled van der Pol oscillators, the characteristic frequencies 
differ twice, and the parameters controlling generation in both oscillators 
undergo a slow periodic counter-phase variation in time. In terms of 
stroboscopic Poincar\'{e} section, the respective four-dimensional mapping 
has a hyperbolic strange attractor of Smale -- Williams type. Qualitative 
reasoning and quantitative data of numerical computations are presented and 
discussed, e.g. Lyapunov exponents and their parameter dependencies. A 
special test for hyperbolicity based on statistical analysis of 
distributions of angles between stable and unstable subspaces of a chaotic 
trajectory has been performed. Perspectives of further comparative studies 
of hyperbolic and non-hyperbolic chaotic dynamics in physical aspect are 
outlined.
\end{abstract}

Mathematical theory of chaotic dynamics based on rigorous axiomatic 
foundation exploits a notion of hyperbolicity, which implies that all 
relevant trajectories in phase space of a dynamical system are of saddle 
type, with well defined stable and unstable directions \cite{1,2,3,4}. Hyperbolic 
systems of dissipative type, contracting the phase space volume, manifest 
robust strange attractors with strong chaotic properties. The robustness 
(structural stability) implies insensitivity of the motions in respect to 
variations of equations governing the dynamics. In particular, positive 
Lyapunov exponent depends on parameters in smooth manner, without flops into 
negative region characteristic to non-hyperbolic attractors. A Cantor-like 
structure of the strange attractor persists without qualitative changes 
(bifurcations), at least while the variations are not too large. Textbook 
examples of these robust strange attractors are represented only by 
artificial mathematical constructions associated with discrete-time models, 
e.g. Plykin attractor and Smale -- Williams attractor (solenoid).

It seems that the mathematical theory of hyperbolic chaos has been never 
applied conclusively to any physical object, although concepts of this 
theory are widely used for interpretation of chaotic behavior of realistic 
nonlinear systems. On the other hand, feasible nonlinear systems with 
complex dynamics, such as Lorenz and R\"{o}ssler equations, chaotic 
self-oscillators, driven nonlinear oscillators etc. do not relate to the 
true hyperbolic class \cite{4,5,6}. As a rule, observable chaos in these systems is 
linked with a so-called quasiattractor, a set in phase space, on which 
chaotic trajectories coexist with stable orbits of high periods (usually, 
they are non-distinguishable in computations at reasonable accuracies). 
Mathematical description of quasiattractors remains a challenging problem, 
although in physical systems the non-hyperbolicity is masked effectively due 
to presence of inevitable noise. In few cases, e.g. in Lorenz model in some 
appropriate domain of parameter space, dynamics is proved to be 
quasi-hyperbolic (with some restrictions concerning violation of smoothness 
conditions) \cite{7}.

I am aware of only two theoretical works, which discuss examples of true 
hyperbolic dynamics in context of systems governed by differential 
equations. One relates to a system called triple linkage, which allows in a 
frictionless case a description in terms of orbits on a surface of negative 
curvature. In dissipative case, it gives rise to a hyperbolic chaotic 
attractor \cite{8}. Another work deals with a 3D flow system motivated by neural 
dynamics and argues in favor of existence of an attractor of Plykin type in 
the Poincar\'{e} map associated with the flow \cite{9}. 

In this Letter, I suggest an essentially simpler and transparent example of 
a non-autonomous flow system, which apparently manifests a hyperbolic 
attractor. In terms of stroboscopic Poincar\'{e} map, it is an attractor of 
the same kind as the Smale -- Williams solenoid, but embedded in a 4D rather 
then 3D state space.

The system is constructed on a basis of two van der Pol oscillators with 
characteristic frequencies $\omega _{0}$ and $2\omega _{0}$, 
respectively. The control parameters of the oscillators responsible for the 
Andronov -- Hopf bifurcations in the autonomous subsystems are forced to 
swing slowly, periodically in time. On a half-period, the first oscillator 
is above the generation threshold, while the second one is below the 
threshold. On another half-period, a situation is opposite. Next, we assume 
that the first oscillator acts on the partner via a quadratic term in the 
equation. The generated second harmonic component serves as a primer for the 
second oscillator, as it comes off the under-threshold state. In turn, the 
second oscillator acts on the first one via a term represented by a product 
of the dynamical variable and an auxiliary signal of frequency 
$\omega_{0}$. Thus, a component with the difference frequency appears, which fits 
resonance range for the first oscillator and serves as a primer as it starts 
to generate. 

Summarizing this description, we write down the following equations:
\begin{equation}
\label{eq1}
\begin{array}{l}
 \ddot {x} - (A\sin {2\pi t /T} - x^2)\dot {x} + \omega _0^2 x = 
\varepsilon y\sin \omega _0 t, \\ 
 \ddot {y} - (- A\sin {2\pi t /T} - y^2)\dot {y} + 4\omega _0^2 y = 
\varepsilon x^2, \\ 
 \end{array}
\end{equation}

\noindent
where $x$ and $y$ are dynamical variables of the first and the second oscillators, 
respectively, $A$ is a constant designating amplitude of the slow swing of the 
control parameters, $\varepsilon $ is a coupling parameter. 

We assume that the period of swing $T$ contains an integer number of periods of 
the auxiliary signal: $T = 2\pi N /\omega_0$. Thus, our 
set of non-autonomous equations has periodic rather than quasiperiodic 
coefficients. It is appropriate to treat the dynamics in terms of 
stroboscopic Poincar\'{e} section using a period-$T$ sequence of time instants. 
The Poincar\'{e} map is four-dimensional and acts in a space of vectors 
$\left\{ {x,\,\,\dot {x}/{\omega _0 },\,y,\,\,\dot {y}/{(2\omega _0 )}} \right\}$. 

The system (\ref{eq1}) operates as follows. Let the first oscillator have some phase 
$\varphi $ on a stage of generation: 
$x \propto \cos (\omega _0 t + \varphi)$. 
Squared value $x^{2}$ contains the second harmonic: $\cos (2\omega _0 t + 
2\varphi )$, and its phase is $2\varphi $. As the half-period comes to the 
end, and the second oscillator becomes excited, the induced oscillations of 
the variable $y$ get the same phase $2 \varphi $. Mixture of these oscillations 
with the auxiliary signal transfers the doubled phase into the original 
frequency range. Hence, on the next stage of excitation the first oscillator 
accepts this phase $2 \varphi $ too. Obviously, on subsequent stages of swing 
the phases of the first oscillator follow approximately the mapping 

\begin{equation}
\label{eq2}
\varphi _{n + 1} = 2\varphi _n \,\,\,(\bmod 2\pi ).
\end{equation}

Figure 1 shows a typical pattern of time dependence of $x$ and $y$ from numerical 
solution of Eqs. (\ref{eq1}) by Runge -- Kutta method for particular parameter 
values $\omega _0 = 2\pi ,\,\,T = N = 10,\,\,A = 3,\,\,\varepsilon = 0.5$ 
together with a diagram of empirical mapping for phase $\varphi _{n + 1} $ 
versus $\varphi _n $. The phases are determined at the centers of the 
excitation stages for the first oscillator:

\begin{equation}
\label{eq2a}
\varphi = \left\{ {{\begin{array}{*{20}l}
{\arctan {(\omega_0^{ - 1} \dot {x}/x)},\,\, x > 0,} \hfill \\
{\arctan {(\omega_0^{ - 1} \dot {x}/x)}+\pi,\,\, x > 0,} \hfill \\
\end{array} }} \right.
\end{equation}

\noindent
and are plotted over a sufficiently large number of the basic periods $T$. The 
mapping for the phase looks, as expected, topologically equivalent to the 
relation (\ref{eq2}). (Some distortions arise due to imperfection of the above 
qualitative considerations and of the definition of phase; the 
correspondence becomes better at larger period ratios $N$.) Chaotic nature of 
the dynamics reveals itself in a random walk of humps in respect to the 
envelope of the generated signal on subsequent periods of swing. 

\begin{figure}[htbp]
\begin{center}
\includegraphics[width=6.5in]{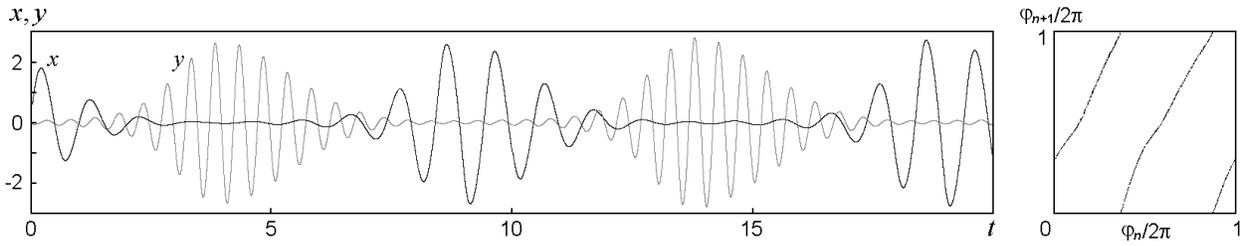}
\end{center}
\label{fig1}
\caption{A typical pattern of time dependence for variables $x$ and $y$ 
obtained from numerical solution of Eqs. (\ref{eq1}) for 
$\omega _0 = 2\pi ,\,\,T = N = 10,\,\,A = 3,\,\,\varepsilon = 0.5$ (a) 
and a diagram of empirical mapping for phases of the first oscillator 
defined in the centers of the stages of excitation 
numbered by $n$}
\end{figure}

In terms of stroboscopic Poincar\'{e} map, attractor of the system 
corresponds exactly to the construction of Smale and Williams. In the 
four-dimensional state space, the direction associated with the phase 
$\varphi $ is expanding and gives rise to Lyapunov exponent estimated as 
$\Lambda _1 \approx T^{ - 1}\log 2$. Three rest directions are contracting, 
and they correspond to a three-dimensional stable manifold of the attractor. 
Three respective Lyapunov exponents are negative. Interpreting the 
stroboscopic Poincar\'{e} mapping, we may imagine a solid torus embedded in 
4-dimensional space and associate one iteration of the map with longitudinal 
stretch of the torus, with contraction in the transversal directions, and 
insertion of the doubly folded ``tube'' into the original torus interior. 

In computations, the Lyapunov exponents were evaluated with a help of 
Benettin's algorithm \cite{10,11} from simultaneous solution of Eqs. (\ref{eq1}) together 
with a collection of four exemplars of the linearized equations for 
perturbations:

\begin{equation}
\label{eq3}
\begin{array}{l}
\ddot {\tilde {x}} + 2x\dot {x}\tilde {x} - (A\sin {2 \pi t /T} - x^2)\dot {\tilde {x}} 
+ \omega _0^2 \tilde{x} = \varepsilon \tilde {y}\sin \omega _0 t, \\ 
\ddot {\tilde {y}} + 2y\dot {y}\tilde {y} - (- A\sin {2 \pi t /T} - y^2)\dot {\tilde {y}} 
+ 4\omega _0^2 \tilde{y} = 2\varepsilon x\tilde {x}. \\ 
\end{array}
\end{equation}

In a course of the solution, at each step of the integration schema, the 
Gram -- Schmidt orthogonalization and normalization were performed for four 
vectors $\left\{ {\tilde {x}(t),{\dot {\tilde {x}}(t)}/{\omega _0 },
\tilde {y}(t),{\dot {\tilde {y}}(t)}/{2\omega _0 }} \right\}$, and the mean rates of 
growth or decrease of the accumulated sums of logarithms of the norms (after 
the orthogonalization but before the normalization) were estimated. As 
found, the Lyapunov exponents for the attractor at the above mentioned 
parameters are 
$\Lambda _1 \approx 0.068 \approx T^{ - 1}\log 2$, $\Lambda_2 \approx - 0.35$, 
$\Lambda _3 \approx - 0.59$, $\Lambda _4 \approx -0.81$. 

If the attractor is indeed hyperbolic, the chaotic dynamics must be robust 
and retain its character under (at least small) variations of the equations. 
As checked, this is indeed the case. In particular, the largest Lyapunov 
exponent is almost independent on parameters, and the rest of them manifest 
regular parameter dependences, as seen in Fig. 2. The left edge of the 
diagram corresponds to violation of the hyperbolicity. 

Dynamical behavior of the same kind is observed at other integer period 
ratios $N$, including essentially smaller ones, e.g. $N$=4. Figure 3 shows 
portrait of the strange attractor in the Poincar\'{e} section in projection 
onto the plane $(x,\dot {x})$ at $\omega _0 = 2\pi ,\,\,T = N = 4,\,\,A = 
8,\,\,\varepsilon = 0.5$. It looks precisely as the Smale -- Williams 
attractor should look like. Observe fractal transversal structure of 
``strips'' constituting the attractor. For this attractor the Lyapunov 
exponents are $\Lambda _1 \approx 0.167$, $\Lambda _2 \approx - 0.72$, 
$\Lambda _3 \approx - 1.03$, $\Lambda _4 \approx - 1.50$. An estimate for 
fractal dimension from Kaplan -- Yorke formula yields $D \approx 1.23$, and 
that from the Grassberger -- Procaccia algorithm is $D \approx 1.26$.

\begin{figure}[htbp]
\begin{center}
\includegraphics[width=4.2in]{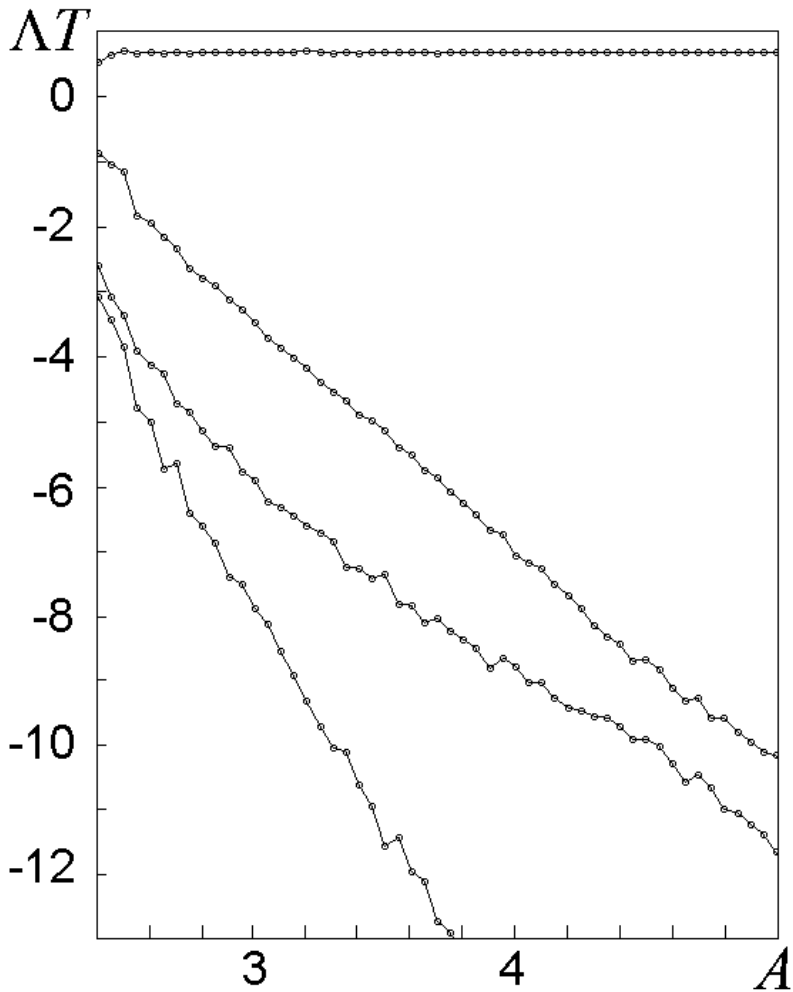}
\end{center}
\label{fig2}
\caption{Computed Lyapunov exponents of the system (\ref{eq1}) versus 
parameter $A$ at $\omega _0 = 2\pi $, $N = T = 10$, $\varepsilon = 0.5$. 
Observe that the largest exponent remains almost constant in the whole 
interval of hyperbolicity being in good agreement with the estimate 
$\Lambda _1 = T^{ - 1}\log 2$. 
}
\end{figure}

It is interesting to perform a direct numerical test for hyperbolicity of 
the attractor. Idea of such test was suggested in Refs. \cite{12} and \cite{13} and 
applied for verification of hyperbolicity of trajectories of dynamical 
systems, which have one stable and one unstable directions. The procedure 
consists in computation of vectors of small perturbations along the 
trajectory in forward and inverse time with measuring angles between the 
forward-time and backward-time vectors at points of the trajectory. If zero 
values of the angle do not occur, i.e. the statistical distribution of the 
angles is essentially separated from zero, one concludes that the dynamics 
is hyperbolic. If the statistical distribution shows non-vanishing 
probability for zero angle, it implies non-hyperbolic behavior because of 
presence of the homoclinic tangencies of stable and unstable manifolds. In 
dissipative case these tangencies are responsible for the occurrence of 
quasiattractor. 

\begin{figure}[htbp]
\begin{center}
\includegraphics[width=4in]{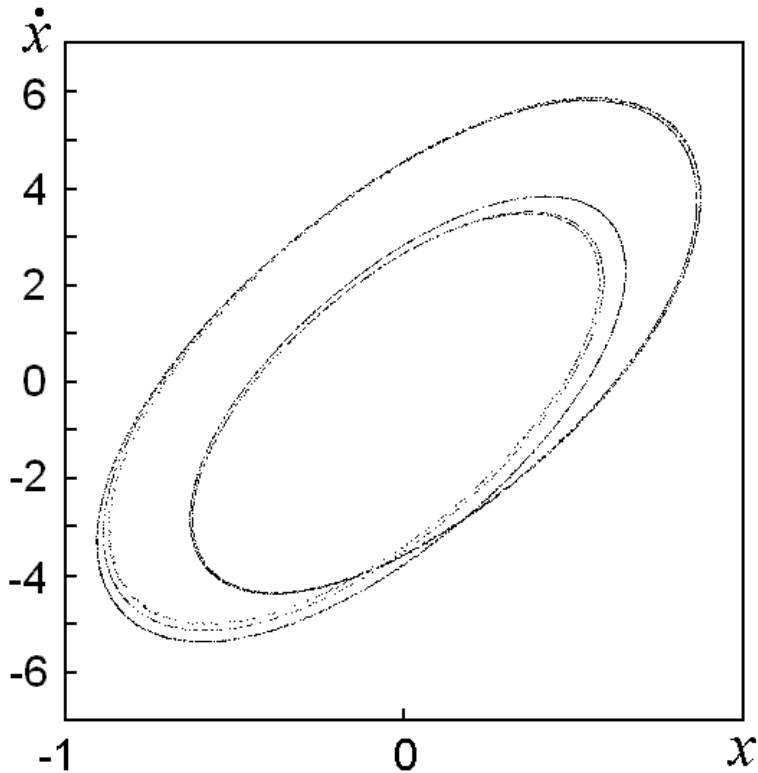}
\end{center}
\label{fig3}
\caption{
Portrait of the strange attractor in the stroboscopic Poincar\'{e} section 
$t = 1\,\,(\bmod N)$ in projection onto the plane 
$(x,\dot {x})$ at $\omega _0 = 2\pi ,\,\,T = N = 4,\,\,A = 8,\,\,\varepsilon = 0.5$
}
\end{figure}

In our system (\ref{eq1}), only unstable subspace is one-dimensional, and the stable 
one is three-dimensional. Therefore, the method needs a modification. An 
adopted algorithm consists in the following. First, we generate a 
sufficiently long representative orbit on the attractor 
$\left\{ {x(t),{\dot {x}(t)}/{\omega _0 },y(t),
{\dot {y}(t)}/{2\omega _0 }} \right\}$ from the numerical solution of Eqs. (\ref{eq1}). 
Then, we solve numerically the equations (\ref{eq3}) for a perturbation forward 
in time. In a course of the solution, normalization of the vector 
${\rm {\bf a}}(t) = \left\{ {\tilde {x}(t),{\dot {\tilde {x}}(t)}/{\omega _0 },
\tilde {y}(t),{\dot {\tilde {y}}(t)}/{2\omega _0 }} \right\}$ is performed after each 
step of integration to exclude the divergence. This vector determines an 
unstable direction at each point of the orbit. Next, we solve a collection 
of three exemplars of equations (\ref{eq3}) in backward time along the same 
trajectory $\left\{ {x(t),{\dot {x}(t)}/{\omega _0},y(t),{\dot {y}(t)}/{2\omega _0 }} \right\}$ 
to get three 
vectors $\left\{ {{\rm {\bf b}}(t),{\rm {\bf c}}(t),{\rm {\bf d}}(t)} 
\right\}$. To avoid dominance of one eigenvector and divergence, we use the 
Gram -- Schmidt orthogonalization and normalization of the vectors at each 
step of the numerical integration. Now, at each point of the trajectory, all 
possible linear combinations of $\left\{ {{\rm {\bf b}}(t),{\rm {\bf 
c}}(t),{\rm {\bf d}}(t)} \right\}$ define a three-dimensional stable 
subspace of perturbation vectors. 

To estimate an angle $\alpha $ between the one-dimensional unstable subspace 
and the three-dimensional stable subspace we first construct a vector ${\rm 
{\bf v}}(t)$ orthogonal to the three-dimensional subspace, with components 
determined from a set of linear equations ${\rm {\bf v}}(t) \cdot {\rm {\bf 
b}}(t) = 0$, ${\rm {\bf v}}(t) \cdot {\rm {\bf c}}(t) = 0$, ${\rm {\bf 
v}}(t) \cdot {\rm {\bf d}}(t) = 0$. Then, we compute an angle 
$\beta  \in [0,\pi/2]$ between the vectors ${\rm {\bf v}}(t)$ and ${\rm {\bf a}}(t)$: 
$\cos \beta = {\left| {{\rm {\bf v}}(t) \cdot {\rm {\bf c}}(t)} \right|}/ 
{\vert {\rm {\bf v}}(t)\vert \vert {\rm {\bf c}}(t)\vert }$ 
and set $\alpha = \beta - \pi/2$. 

\begin{figure}[htbp]
\begin{center}
\includegraphics[width=6.5in]{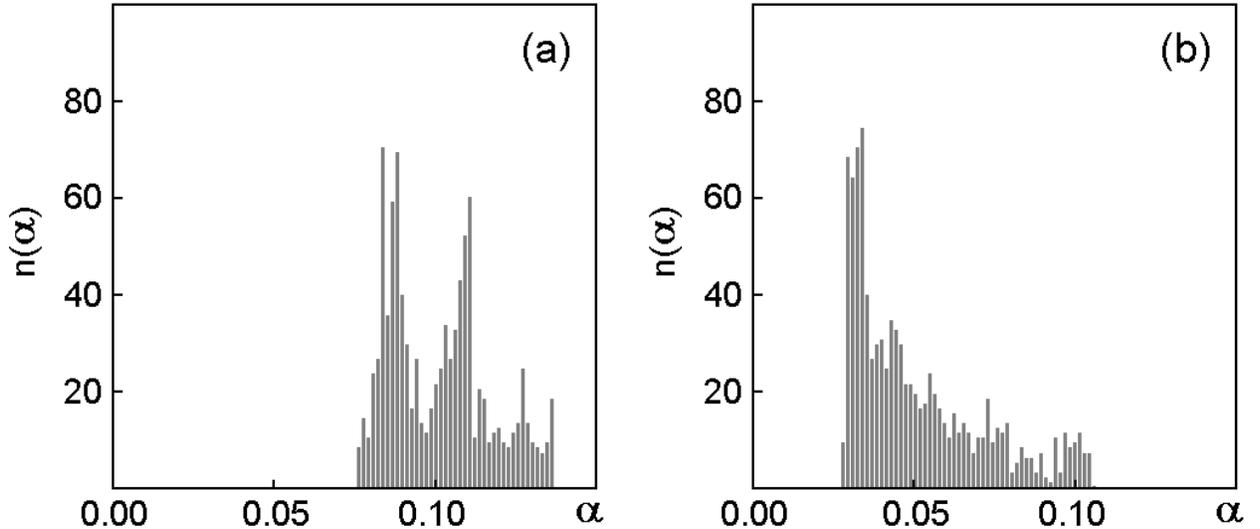}
\end{center}
\label{fig4}
\caption{Histograms for distributions of angles $\alpha $ between 
the stable and unstable subspaces for the system (\ref{eq1}) with $\omega _0 = 2\pi 
,\,\,\,\varepsilon = 0.5$ obtained from computational procedure described in 
the text: (a)$N = 10,\,\,A = 3$ and (b) $N = 4,\,\,A = 8$}
\end{figure}

Figure 4 shows histogram for the distribution of angles $\alpha $ between 
the stable and unstable subspaces for the system (\ref{eq1}) obtained from 
computations at the two mentioned sets of parameter values. Observe clearly 
visible separation of the distributions from zero values of $\alpha $. So, 
the test confirms hyperbolicity of the attractors. 

In spite of simplicity of the presented example, I believe it is significant 
as a feasible system, which may be designed as a physical device, e.g. on a 
basis of two interacting electronic oscillators. It opens an opportunity for 
experimental studies of hyperbolic chaos and its features predicted by the 
mathematical hyperbolic theory (robustness, continuity of the invariant 
measure, insensitivity of statistical characteristics of the motions in 
respect to noise, etc.). In addition, it makes conclusive comparative 
examination of dynamics of hyperbolic and non-hyperbolic systems. 

In a sense, breakthrough into the hyperbolic domain is a decisive step. Now 
one can construct many other examples of systems with hyperbolic attractors: 
Because of robustness of such attractor, any variation of 
the right-hand parts of the equations will not destroy the hyperbolicity, at 
least while they are not too large. Apparently, in this way it is possible to design 
examples of autonomous systems with hyperbolic strange 
attractorsb as well, via modification of the system supplimenting  
additional equations for dynamical variables, which would represent the swing and the 
auxiliary signals. 

The author thanks A. Pikovsky and M. Rosenblum for helpful discussion. The 
work has been performed under partial support from RFBR (grant No 03-02-16192) and
from CRDF via the Research Educational Center of
Saratov University (Grant No. REC-006).

\end{document}